\documentstyle[12pt,aasms4,psfig]{article}
\evensidemargin=\oddsidemargin
\renewcommand{\baselinestretch}{2}
\def\ltsima{$\; \buildrel < \over \sim \;$}
\def\lsim{\lower.5ex\hbox{\ltsima}}
\def\gtsima{$\; \buildrel > \over \sim \;$}
\def\gsim{\lower.5ex\hbox{\gtsima}}

\newcommand{\be}{\begin{equation}}
\newcommand{\en}{\end{equation}}

\def\cmdue {\rm \ cm^{-2}}

\begin{document}
\begin{center}

\vspace{1cm}

{\bf 
The shock break-out of GRB060218/SN2006aj
}

{S. Campana$^1$, V. Mangano$^2$, A. J. Blustin$^3$, P. Brown$^4$,
D. N. Burrows$^4$, G. Chincarini$^{5,1}$, J. R. Cummings$^{6,7}$, G. Cusumano$^2$,
M. Della Valle$^{8,9}$, D. Malesani$^{10}$, P. M\'esz\'aros$^{4,11}$, J. A. Nousek$^{4}$,
M. Page$^3$, T. Sakamoto$^{6,7}$, E. Waxman$^{12}$, B. Zhang$^{13}$,
Z. G. Dai$^{14,13}$, N. Gehrels$^{6}$, S. Immler$^6$, F. E. Marshall$^{6}$,
K. O. Mason$^{15}$, A. Moretti$^1$, P. T. O'Brien$^{16}$, J. P. Osborne$^{16}$,
K. L. Page$^{16}$, P. Romano$^{1}$, P. W. A. Roming$^{4}$, G. Tagliaferri$^{1}$, 
L.R. Cominsky$^{17}$, P. Giommi$^{18}$, O. Godet$^{16}$, J. A. Kennea$^4$,
H. Krimm$^{6,19}$, L. Angelini$^{6}$, S. D. Barthelmy$^{6}$,
P. T. Boyd$^{6}$, D. M. Palmer$^{20}$, A. A. Wells$^{16}$, N. E. White$^{6}$
}

\end{center}

\noindent $^1$ INAF - Osservatorio Astronomico di Brera, via E. Bianchi 46, I-23807
Merate (LC), Italy.

\noindent $^2$ INAF - Istituto di Astrofisica Spaziale e Fisica Cosmica di Palermo, via U. La
Malfa 153, I-90146 Palermo, Italy.

\noindent $^3$ UCL Mullard Space Science Laboratory, Holmbury St. Mary,
Dorking, Surrey RH5 6NT, UK.

\noindent $^4$ Department of Astronomy and Astrophysics, Pennsylvania State
University, University Park, Pennsylvania 16802, USA. 
 
\noindent $^5$ Universit\`a degli studi di Milano Bicocca, piazza delle
Scienze 3, I-20126 Milano, Italy. 

\noindent $^6$ NASA - Goddard Space Flight Center, Greenbelt, Maryland 20771, USA. 

\noindent $^7$ National Research Council, 2101 Constitution Ave NW, Washington
DC 20418, USA. 

\noindent $^8$ INAF - Osservatorio Astrofisico di Arcetri, largo E. Fermi 5,
I-50125 Firenze, Italy.

\noindent $^9$ Kavli Institute for Theoretical Physics, UC Santa Barbara,
California 93106, USA.

\noindent $^{10}$ International School for Advanced Studies (SISSA-ISAS), via
Beirut 2-4, I-34014 Trieste, Italy. 

\noindent $^{11}$ Department of Physics, Pennsylvania State
University, University Park, Pennsylvania 16802, USA.
 
\noindent $^{12}$ Physics Faculty, Weizmann Institute,
Rehovot 76100, Israel.

\noindent $^{13}$ Department of Physics, University of Nevada, Box 454002, Las
Vegas, Nevada 89154-4002, USA.

\noindent $^{14}$ Department of Astronomy, Nanjing University, Nanjing 210093,
China. 

\noindent $^{15}$ PPARC, Polaris House, North Star Avenue, Swindon SN2 1SZ, UK.

\noindent $^{16}$ Department of Physics and Astronomy, University of
Leicester, University Road, Leicester LE1 7RH, UK. 

\noindent $^{17}$ Department of Physics and Astronomy, Sonoma State
University, Rohnert Park, California 94928-3609, USA.

\noindent $^{18}$ ASI Science Data Center, via G. Galilei, I-00044 Frascati
(Roma), Italy. 

\noindent $^{19}$ Universities Space Research Association, 10211 Wincopin
Circle Suite 500, Columbia, Maryland 21044-3431, USA.

\noindent $^{20}$ Los Alamos National Laboratory, P.O. Box 1663, Los Alamos,
New Mexico 87545, USA.

\newpage

{\bf 
Although the link between long Gamma Ray Bursts (GRBs) and supernovae (SNe)
has been established$^{1-4}$,
hitherto there have been no observations of the beginning of a 
supernova explosion and its intimate link to a GRB. 
In particular, we do not know however how a GRB jet emerges from the star surface nor
how a GRB progenitor explodes.
Here we report on 
observations of the close GRB060218$^{5}$ 
and its connection to SN2006aj$^{6}$.
In addition to the classical non-thermal emission, GRB060218 shows a
thermal
component in its X--ray spectrum, which cools and 
shifts into the optical/UV band as time passes. 
We interpret these features as arising from the break out of a shock driven by
a mildly relativistic shell into the dense wind surrounding the
progenitor$^{7}$. 
Our observations 
allow us for the
first time to catch a SN in the act of exploding, 
to directly observe the shock break-out
and to provide strong evidence that the GRB progenitor was a Wolf-Rayet star.  
}

GRB060218 was detected with the BAT instrument$^{8}$
onboard Swift$^{9}$ 
on 2006 February 18.149 UT$^{5}$.
The burst profile is unusually long with a $T_{90}$ (the time
interval containing $90\%$ of the flux) of $2100\pm100$ s (Fig. 1). The
flux slowly rose to the peak at $431\pm60$ s ($90\%$ containment, times are
measured from the BAT trigger time). 
Swift slewed autonomously to the newly discovered burst.
The X--ray
Telescope (XRT$^{12}$) found a bright source,
which rose smoothly to a peak of $\sim 100$ counts s$^{-1}$ (0.3--10 keV) at $985\pm15$ s. 
The X--ray flux then decayed exponentially 
with an $e-$folding time of $2100\pm50$ s, 
followed around 10 ks by a shallower power-law decay similar to that seen in
typical GRB afterglows$^{14-15}$
(Fig. 2).  
The UltraViolet/Optical Telescope (UVOT$^{13}$)
found emission steadily brightening by a factor of 5--10 after the
first detection, peaking in a broad plateau first in the UV ($31.3\pm1.8$ ks at
188 nm) and later in the optical ($39.6\pm2.5$ ks at 439 nm). The light
curves reached a minimum at about 200 ks, after which the UV light curves
remained constant while a rebrightening is seen in the optical bands, peaking
again at about 700--800 ks (Fig. 2).  

Soon after the Swift discovery, low-resolution spectra of the optical
afterglow and host galaxy revealed strong emission lines at a redshift of
$z=0.033$ (ref. 16). 
Spectroscopic indications of the presence of a rising supernova (designated
SN2006aj) were found three days after the burst$^{6,17}$  
with broad emission features consistent with
a type Ic SN (due to a lack of hydrogen and helium lines).

The Swift instruments provided valuable spectral information. The high energy
spectra soften with time and can be fit with (cut-off) power-laws.
This power-law component can be ascribed to the usual GRB jet and afterglow.
The most striking feature, however, is the presence of a soft component in the
X--ray spectrum,
that is present in the XRT 
up to $\sim 10,000$ s. 
The blackbody component shows a marginally decreasing temperature ($k\,T\sim
0.17$~keV, where $k$ is the Boltzmann
constant), and a clear increase in luminosity with
time, corresponding to an increase in the apparent emission radius from
$R^{\rm X}_{\rm BB}=(5.2\pm0.5)\times10^{11}$ cm to $R^{\rm X}_{\rm 
BB}=(1.2\pm0.1)\times 10^{12}$~cm (Fig.~3).
During the rapid decay ($t\sim 7000$ s), a blackbody component is still
present in the data with a marginally cooler temperature ($k\,T=0.10\pm0.05$
keV) and a comparable emission radius [$R^{\rm X}_{\rm BB}=(6.5^{+14}_{-4.4})\times
10^{11}$ cm].  
In the optical/UV band at 9 hours (32 ks) 
the blackbody peak
is still above the UVOT energy range.
At 120 ks 
the peak of the blackbody emission is 
within the UVOT passband, and the inferred temperature and radius are
$k\,T=3.7^{+1.9}_{-0.9}$ eV and $R^{\rm UV}_{\rm BB}=3.29^{+0.94}_{-0.93}\times 10^{14}$ cm, 
implying an expansion speed of $(2.7\pm0.8)\times 10^9$ cm s$^{-1}$.   
This estimate 
is consistent with what we would expect for a SN 
and it is also consistent with the line broadening observed in the optical
spectra. 

The thermal components are the key to interpreting this anomalous GRB. 
The high temperature (two million degrees) of the thermal X--ray component suggests
that the radiation is emitted by a shock-heated plasma. 
The characteristic radius
of the emitting region, $R_{\rm
shell}\sim(E/a\,T^4)^{1/3}\sim5\times10^{12}$~cm ($E$ is the GRB isotropic
energy and
$a$ is the radiation density constant), 
corresponds to the radius of a blue supergiant progenitor. However, the lack of 
hydrogen lines in the SN spectrum suggests a much more compact source.
The large emission radius may 
be explained in this case by the existence of a massive stellar wind
surrounding the progenitor, as is common for Wolf-Rayet stars. The thermal
radiation is observed once the shock driven into the wind reaches a radius, 
$\sim R_{\rm shell}$, where the wind becomes optically thin. 

The characteristic variability time is $R_{\rm shell}/c\sim 200$~s, consistent
with the smoothness of the X--ray pulse and the rapid thermal X--ray flux
decrease at the end of the pulse. We interpret this as providing, for 
the first time, a direct measurement of the shock break-out$^{19,20}$ of the stellar 
envelope and the stellar wind (first investigated by Colgate$^{21}$). 
The fact that $R_{\rm shell}$ is larger than
$R_{BB}^{X}$ suggests that the shock expands in a non-spherical manner, reaching
different points on the $R_{\rm shell}$ sphere at different times. This
may be due to a non-spherical explosion (e.g. the presence of a jet), or a
non-spherical wind$^{22,23}$. 
In addition, the shock break-out interpretation provides us 
with a delay between the SN explosion and the GRB start to be $\lsim 4$
ks$^{24}$ (Fig. 1).

As the shock propagates into the wind, it compresses the wind plasma into a
thin shell.  
The mass of this shell may be inferred from
the requirement that its optical depth be close to unity, 
$M_{\rm shell}\approx 4\,\pi\, R^2_{\rm shell}/\kappa\approx5\times10^{-7}\,M_\odot$
($\kappa\approx0.34\,{\rm cm^2\, g^{-1}}$ is the opacity). This implies that the 
wind mass-loss rate is $\dot{M}\approx M_{\rm shell}\,v_{\rm wind}/R_{\rm shell}
\approx 3\times10^{-4}\,M_\odot\, {\rm yr^{-1}}$, for a wind velocity $v_{\rm
wind}=10^8 \, {\rm cm\, s^{-1}}$,
typical for Wolf-Rayet stars. Since the thermal energy density behind a
radiation-dominated shock is $a\,T^4\sim 3\,\rho\,v_s^2$ ($\rho$ is the wind
density at $R_{\rm shell}$ and $v_s$ the shock velocity) we have
$\rho\sim10^{-12}\,{\rm g \, cm^{-3}}$, which
implies that the shock must be (mildly) relativistic, $v_s\simeq c$. This is 
similar to GRB980425/SN1998bw, where the ejection of a mildly
relativistic shell with energy of $\simeq 5\times 10^{49}$~erg is believed to have
powered radio$^{25-27}$
and X--ray emission$^{7}$.

The optical-UV emission observed at early time, $t\lsim 10^4$~s, may be accounted for as
the low-energy tail of the thermal X--ray emission produced by the (radiation)
shock driven into the wind. At later time, the optical-UV emission is well
above that expected from the (collisionless) shock driven into the wind. This
emission is most likely due to the expanding envelope of the star, which was
heated by the shock passage to a much higher temperature. Initially, this envelope is
hidden by the wind. As the star and wind expand, the photosphere propagates
inward,  revealing shocked stellar plasma. As the star expands,  
the radiation temperature decreases and the apparent radius increases (Fig. 3). 
The radius inferred at the peak of the UV emission, $R^{\rm UV}_{\rm BB}\sim
3\times10^{14}$~cm, implies that emission is arising from the outer  
$\sim 4\,\pi\, (R^{\rm UV}_{\rm BB})^2/\kappa\sim 10^{-3}~M_\odot$ shell at the edge of
the shocked star. As the photosphere rapidly cools, this component of the
emission fades. 
The UV light continues to plummet as cooler temperatures allow elements to
recombine and line blanketing to set in, while radioactive decay causes the
optical light to begin rising to the primary maximum normally seen in SN light
curves (Fig. 2).

Since the wind shell is clearly larger than the progenitor radius, we infer
that the star radius is definitely smaller than $5\times
10^{12}$ cm. Assuming a linear expansion at the beginning (due to light travel
time effects) one can estimate a star radius of $R_{\rm star}\sim (4\pm1)\times 10^{11}$
cm. This is smaller than the radius of the progenitors of type II SNe, like
blue supergiants ($4\times10^{12}$ cm for SN1987A, ref. 28)
or red supergiants ($3\times10^{13}$ cm). Our results
unambiguously indicate that the progenitor of GRB060218/SN2006aj 
was a compact massive star, most likely a Wolf-Rayet star.

\newpage

\references

\noindent 1. Woosley, S. E., Gamma--ray bursts from stellar mass accretion disks around black holes. 
Astrophys. J., {\bf 405}, 273--277 (1993).

\noindent 2. Paczy\'nski, B., Are Gamma-Ray Bursts in Star-Forming Regions? Astrophys. J., {\bf 494}, 
L45--L48 (1993).

\noindent 3. MacFadyen, A. I., Woosley, S. E., Collapsars: Gamma-Ray Bursts and
Explosions in ``Failed Supernovae'', Astrophys. J., {\bf 524}, 262--289 (1999).

\noindent 4. Galama, T. J., {\it et al.}, An unusual supernova in the error box of the gamma-ray burst 
of 25 April 1998. Nature, {\bf 395}, 670--672 (1998).

\noindent 5. Cusumano, G., {\it et al.}, GRB060218: Swift-BAT detection of a possible burst, 
GCN 4775 (2006).

\noindent 6. Masetti, N., {\it et al.}, GRB060218: VLT spectroscopy. GCN 4803 (2006). 

\noindent 7. Waxman, E., Does the Detection of X--Ray Emission from SN1998bw Support Its 
Association with GRB980425? Astrophys. J., {\bf 605}, L97--L100 (2004).

\noindent 8. Barthelmy, S. D., {\it et al.}, The Burst Alert Telescope (BAT) on the SWIFT Midex Mission
Sp. Sci. Rev., {\bf 120}, 143--164 (2005).

\noindent 9. Gehrels, N., {\it et al.}, The Swift gamma ray burst mission}.  Astrophys. J. 
{\bf 611}, 1005--1020 (2004).

\noindent 10. Amati, L., {\it et al.}, GRB060218: E$_{\rm p,i}$ - E$_{\rm iso}$ correlation. GCN 4846 (2006).

\noindent 11. Heise, J., {\it et al.}, X--Ray Flashes and X--Ray Rich Gamma
Ray Bursts. Gamma-Ray Bursts in the Afterglow Era, Proceedings of ``Gamma-Ray
Bursts in the Afterglow Era'', E. Costa, F. Frontera, J. Hjorth eds., (Berlin
Heidelberg: Springer), p. 16--21 (2001). 

\noindent 12. Burrows, D. N., {\it et al.}, The Swift X--Ray Telescope. Sp. Sci. Rev., {\bf 120}, 165--195 (2005).

\noindent 13. Roming, P. W. A., {\it et al.}, The Swift Ultra-Violet/Optical Telescope. Sp. Sci. Rev. {\bf 120}, 95--142 (2005).

\noindent 14. Tagliaferri, G.,  {\it et al.}, An unexpectedly rapid decline in the X--ray afterglow 
emission of long $\gamma$-ray bursts. Nature, {\bf 436}, 985--988 (2005).

\noindent 15. O'Brien, P. T., {\it et al.}, The early X--ray emission from GRBs. Astrophys. J. submitted (astro-ph/0601125).

\noindent 16. Mirabal, N., Halpern J.P., GRB060218: MDM Redshift. GCN 4792 (2006).

\noindent 17. Pian, E., {\it et al.}, Gamma-Ray Burst associated Supernovae: Outliers 
become Mainstream. Nature submitted (astro-ph/0603530).

\noindent 18. Pei, Y. C., Interstellar dust from the Milky Way to the Magellanic Clouds.  
Astrophys. J., {\bf 395}, 130--139 (1992).

\noindent 19. Ensman, L., Burrows, A., Shock breakout in SN1987A. 
Astrophys. J., {\bf 393}, 742--755 (1992).

\noindent 20. Tan, J. C., Matzner, C. D.; McKee, C. F., Trans-Relativistic 
Blast Waves in Supernovae as Gamma-Ray Burst Progenitors. Astrophys. J., 
{\bf 551}, 946--972 (2001).

\noindent 21. Colgate, S. A., Prompt gamma rays and X--rays from supernovae. 
Can. J. Phys., {\bf 46}, 476 (1968).

\noindent 22. Mazzali, P. A. {\it et al.}, An asymmetric, energetic type Ic
supernova viewed off-axis and a link to gamma-ray bursts. Science, {\bf 308},
1284--1287 (2005).  

\noindent 23. Leonard, D. C., {\it et al.}, A non-spherical core in the
explosion of supernova SN2004dj. Nature, in the press (astro-ph/0603297).

\noindent 24. Norris, J. P., Bonnell, J. T., How can the SN-GRB time delay be
measured? AIP Conference Proceedings, 727, 412--415 (2004). 

\noindent 25. Kulkarni, S. R., {\it et al.}, Radio emission from the unusual supernova
1998bw and its association with the gamma-ray burst of 25 April 1998. Nature,
{\bf 395}, 663--669 (1998). 

\noindent 26. Waxman, E.,  Loeb, A., A Subrelativistic Shock Model for the
Radio Emission of SN1998bw. Astrophys. J., {\bf 515}, 721--725 (1999).

\noindent 27. Li, Z.-Y., Chevalier, R. A., Radio Supernova SN1998bw and Its Relation to
GRB980425., Astrophys. J., {\bf 526}, 716--726 (1999).

\noindent 28. Arnett, W. D., {\it et al.}, Supernova
1987A. Ann. Rev. Astron. Astrophys., {\bf 27}, 629--700 (1989).

\bigskip

\noindent
{\bf Acknowledgements.} The authors acknowledge support from ASI, NASA and
PPARC.

\bigskip

\noindent
{\bf Competing interests.} The authors declare to have no competing financial interests. 

\bigskip

\noindent 
{\bf Corresponding author:} Sergio Campana, INAF - Osservatorio astronomico di
Brera, Via E. Bianchi 46, I-23807 Merate (LC), Italy, e-mail: sergio.campana@brera.inaf.it

\newpage

\renewcommand{\baselinestretch}{1}

\begin{figure}
\caption[]
{{\bf Early Swift light curve of GRB060218.}
GRB060218 was discovered by the BAT when it came into the BAT field of view
during a pre-planned slew. There is no emission at the GRB location up to
--3509 s.
Swift slewed again to the burst position 
and the XRT and UVOT began observing GRB060218 159 s later. The BAT light
curve is shown with green open squares. The XRT data are shown with black open
circles. For each BAT point we converted the
observed count rate to flux (15--150 keV band) using the observed spectra. 
Combined BAT and XRT spectra were fit with cut-off power-law plus a blackbody, 
absorbed by interstellar matter in our Galaxy [column density of
$(0.9-1.1)\times10^{21}\cmdue$] and in the host galaxy at redshift
$z=0.033$. The host galaxy column density is $N_H^{z}=5.0\times
10^{21}\cmdue$. Errors are at $1\,\sigma$ significance. 
At a redshift $z=0.033$ (corresponding to a distance of 145 Mpc with $H_0=70$
km s$^{-1}$ Mpc$^{-1}$) the isotropic equivalent energy, extrapolated to the
1--10,000 keV rest frame energy band, is $E_{\rm
iso}=(6.2\pm0.3)\times10^{49}$ erg. The peak energy in the GRB spectrum is at $E_{\rm
p}=4.9^{+0.4}_{-0.3}$ keV. These values are consistent with the Amati
correlation, suggesting that GRB060218 is not an off-axis event$^{10}$. 
This conclusion is also supported by the lack of achromatic rise
behavior of the light curve in the three Swift observation bands. The BAT
fluence is dominated by soft X--ray photons and this burst can be classified
as an X--ray flash$^{11}$.\\
A $V$ band light curve is shown with
red filled circles. For clarity the $V$ flux has been multiplied by a factor of 100.
Magnitudes have been converted to fluxes using standard UVOT
zeropoints and multiplying the specific flux by the filter Full Width at Half
Maximum. Gaps in the light curve are due to the automated periodic
change of filters during the first observation of the GRB.\\
\label{fig:light2}}
\centerline{\psfig{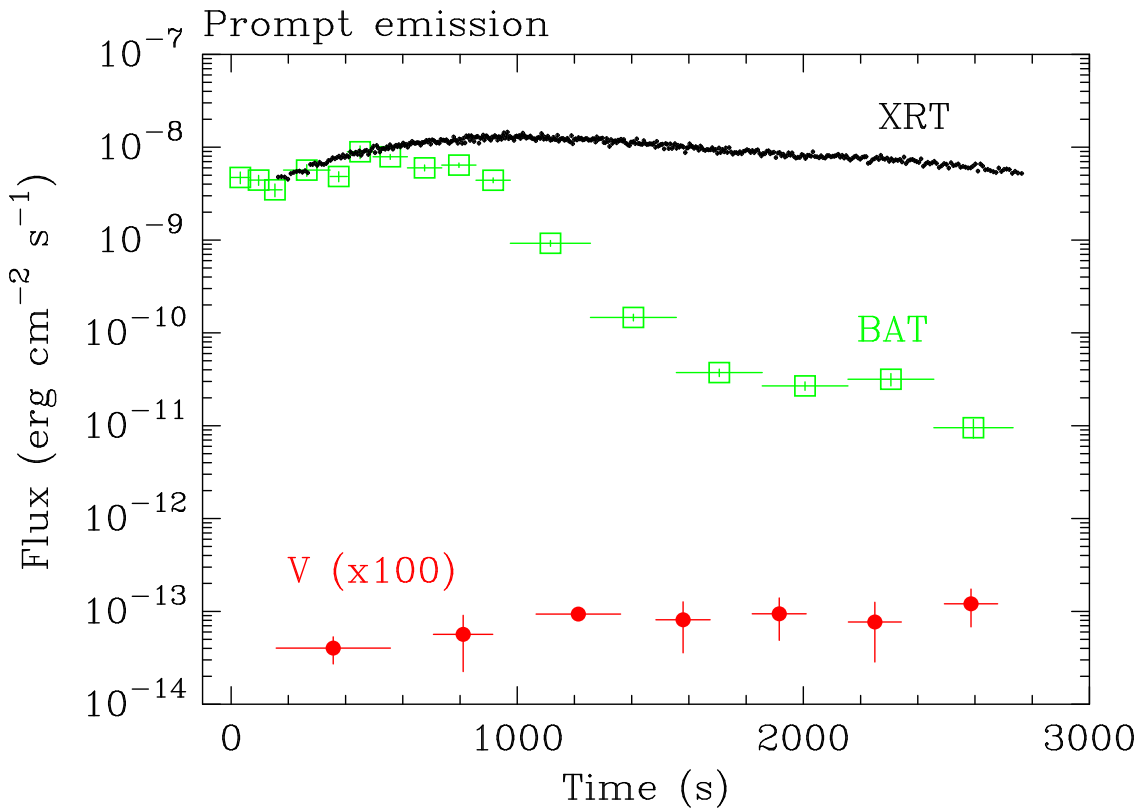}}
\end{figure}  

\newpage 

\begin{figure}[!t]
\caption[]{{\bf Long-term Swift light curve of GRB060218.}
{\it Upper panel}: the XRT light curve (0.3--10 keV) is shown with open black
circles. Count rate-to-flux conversion factors were derived from
time-dependent spectral analysis. We also plot with open black squares the 
contribution to the 0.3--10 keV flux 
by the blackbody component. Its percentage contribution is increasing with
time, becoming dominant at the end of the exponential decay.
The X--ray light curve has a long, slow power-law
rise followed by an exponential (or steep power-law) decay. At about 10,000 s
the light curve breaks to a shallower power-law decay with index $-1.2\pm0.1$
characteristic of typical GRB afterglows.  
This classical afterglow can be naturally accounted for by a shock driven into
the wind by a shell with kinetic energy $E_{\rm shell}\sim10^{49}\ {\rm
erg}$. The $t^{-1}$ flux decline is valid at the stage where the
shell is being decelerated by the wind with the deceleration phase beginning
at $t_{\rm dec}\lsim 10^4$~s for $\dot{M}\gsim 10^{-4}(v_{\rm wind}/10^8\,{\rm
cm\, s^{-1}})\,M_\odot\,{\rm yr^{-1}}$, consistent with the mass-loss rate
inferred from the thermal X--ray component.\\
{\it Lower panel}: the UVOT light curve. Filled circles of different colors
represent different UVOT filters: red -- $V$ (centered at 544 nm); green --
$B$ (439 nm), blue -- $U$ (345 nm), light blue -- UVW1 (251 nm); magenta -- UVM1
(217 nm) and yellow -- UVW2 (188 nm). Specific fluxes have been multiplied by
their FWHM widths (75, 98, 88, 70, 51 and 76 nm, respectively). 
Data have been rebinned to increase the signal to noise ratio. 
The UV band light curve peaks
at about 30 ks due to the shock break-out from the outer stellar surface and
the surrounding dense stellar wind, while the optical band peaks at about 800
ks due to radioactive heating in the SN ejecta.\\
\label{fig:light1}}
\newpage
\centerline{\psfig{figure=2006-03-02579A_f2.ps,width=15truecm,angle=-90}}
\end{figure}  

\newpage

\begin{figure}[!t]
\caption[]{{\bf Evolution of the soft thermal component temperature
and radius.} 
{\it Upper panel}: evolution of the temperature of the soft thermal
component. The joint BAT and XRT spectrum has been fit with a blackbody component plus a
(cut-off) power-law in the first $\sim 3,000$ s (see also
the caption of Fig. 1). The last point (circled in green) comes from 
a fit to the six UVOT filters, assuming a
blackbody model with Galactic reddening [$E(B-V)=0.14$] and host galaxy
reddening. This reddening has been determined by fitting the Rayleigh-Jeans tail of the 
blackbody emission at 32 ks (9 hours). The data require an intrinsic
$E(B-V)=0.20\pm0.03$ (assuming a Small Magellanic Cloud reddening law$^{18}$).
{\it Lower panel}: evolution of the radius of the soft thermal component. The
last point (circled in green) comes from the fitting of UVOT data. 
The continuous line represents a linear fit to the data.\\
\label{fig:light3}
}
\centerline{\psfig{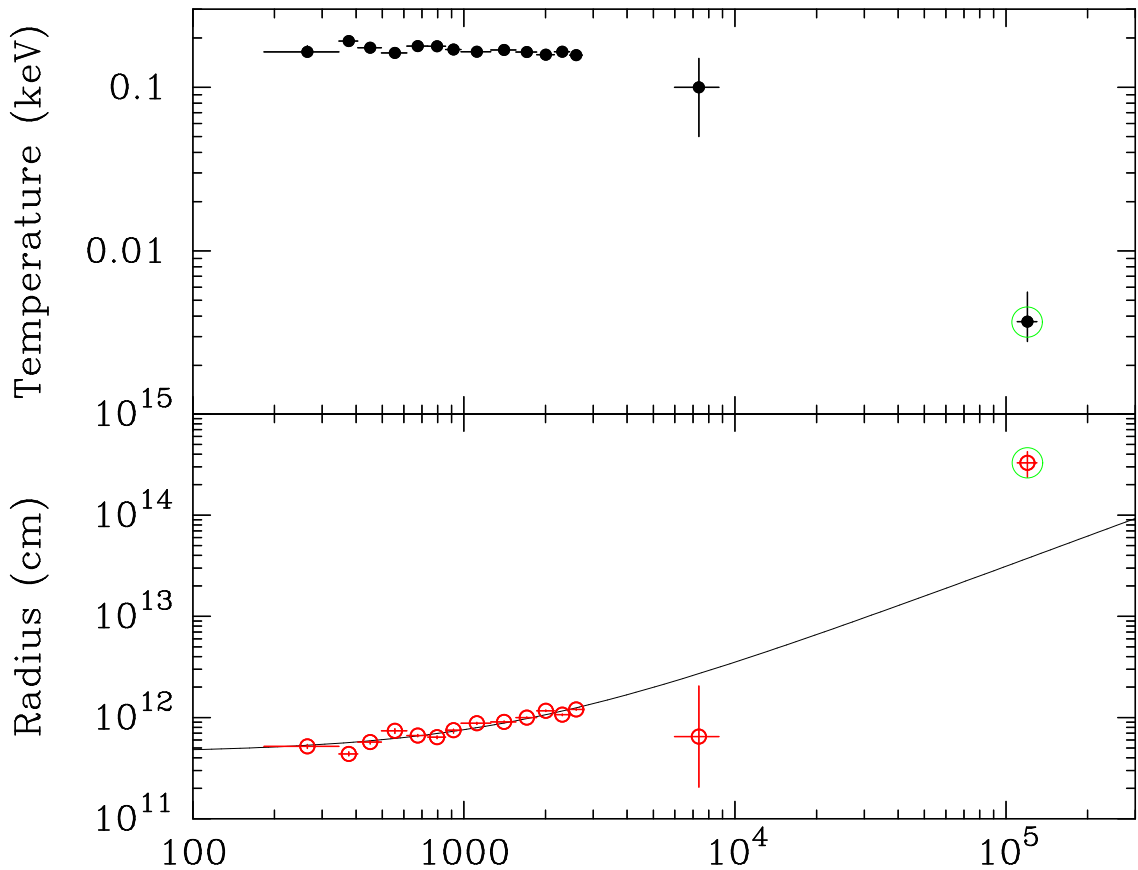}}
\end{figure}

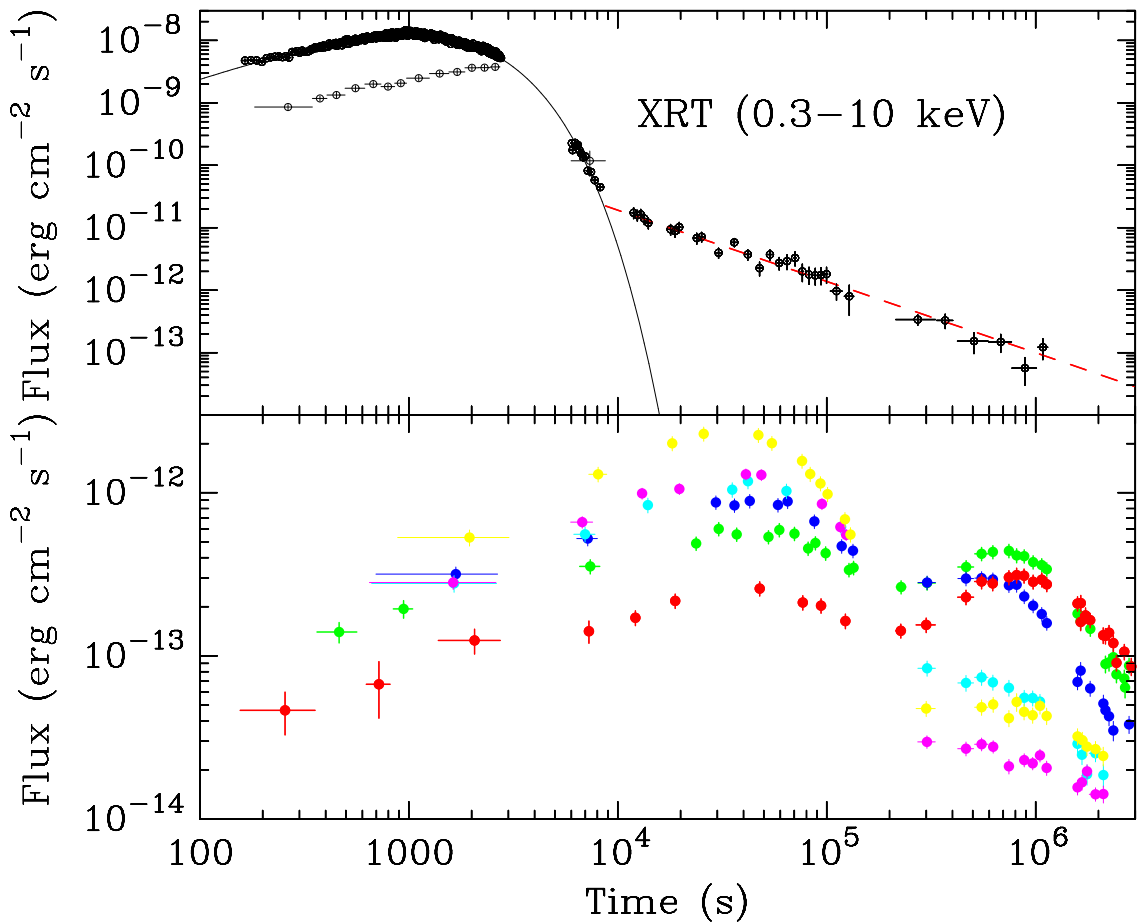
\end{document}